\documentstyle[preprint,aps,prl,amsfonts,]{revtex}
\begin{document}
\preprint{HEP}
\title{Stability of SU(2) Quantum Skyrmion and Static Properties of Nucleons}
\author{A. Acus and E. Norvai\v sas}
\address{Institute of Theoretical Physics and Astronomy\\
Go\v stauto 12, Vilnius 2600, Lithuania }
\date{}
\maketitle

\begin{abstract}
The Skyrme model is considered quantum mechanically {\it ab initio }in 
various irreducible representations of the SU(2) group. The canonical 
quantization procedure yields negative mass correction ensuring existence of 
stabile soliton solution even in chiral limit. The evaluated static 
properties of nucleons (masses, magnetic moments, radii {\it etc.}) are in a 
good agreement with experimental data.
\end{abstract}

\narrowtext
Quantum chromodynamics is the fundamental theory of strong interactions.
However it does not provide a simple description of low-energy hadron 
physics. One of the simplest models which describes the main
features of low-energy physics is the SU(2)
Skyrme model~\cite{skyrme61a}. The topological solutions of the model are 
quantized in collective coordinates approach~\cite{adkins83} and 
interpreted as baryons. The semiclassical quantization procedure leads to 
the energy expression which minimizing does not provide any stabile quantum 
skyrmion ~\cite{bander84,braaten85}. In this letter we make use of  
canonical quantization procedure, developed in~\cite{norvaisas94,acus96a} 
and show for the first time existence of stabile quantum skyrmion 
even in the chiral limit. The numerical solution of equation and predicted 
static nucleon observables are in a good agreement with empirical values. 
The pion mass is also evaluated using nucleon characteristics only.

Usually the Skyrme model is defined in the 
fundamental representation of the SU(2) group. The present analysis is 
based on the chirally symmetric Lagrangian density
\begin{equation}
{\cal L[}U(\bbox{x},t){\cal ]}=-{f_\pi ^2\over 4}{\rm Tr}\{R_\mu R^\mu \}+{
{1\over 32e^2}{\rm Tr}}\{[R_\mu ,R_\nu ]^2\},
\label{f1} 
\end{equation}
where the unitary field $U(\bbox{x},t)=D^j(\bbox{\alpha }(\bbox{x},t))$ 
belongs to an irreducible representation of general dimension of the SU(2) 
group. Here $f_\pi $ (the pion decay constant) and $e$ are parameters 
and $R_\mu =(\partial _\mu U)U^{\dagger }$ is 'right' current. The 
elements of $(2j+1)$ dimension  matrices $D^j$ are the Wigner D-functions 
expressed in terms of three unconstrained Euler angles $\bbox{\alpha 
}=(\alpha ^1,\alpha ^2,\alpha ^3)$~\cite{norvaisas94}. We quantize the 
skyrmion in collective coordinate approach 
\begin{equation}
U({\bf x},{\bf q}(t))=A\left( {\bf q}(t)\right) U_0({\bf 
x})A^{\dagger}\left( {\bf q}(t)\right)\label{f2} 
\end{equation}
{\it ab initio}~\cite{acus96a,fujii87}. The $U_0$ matrix is the 
classical hedgehog ansatz in j representation. The collective coordinates  
${\bf q}(t)$ (the Euler angles in matrix $A$) are dynamical variables  
such that $[\dot q^a,\,q^b]\neq 0$. The energy of quantum skyrmion 
corresponding to Lagrangian density~(\ref{f1}) in representation $j$ with 
spin-isospin $\ell $ and chiral angle $F(r)$, have a form
\begin{equation}
E(j,\ell ,F)=M(F)+\Delta M_j(F)+{\ell (\ell +1)\over 2a(F)},\label{f3} 
\end{equation}
where
\begin{eqnarray}
M(F)={f_\pi \over e}\tilde M(F)=&& 
2\pi {f_\pi \over e}\int {\rm d}\tilde{r}\tilde{r}^2\bigl[ F^{\prime
2}\nonumber\\
&&+{\sin ^2F\over \tilde{r}^2}\bigl( 2+2F^{\prime 2}
+{\sin ^2F\over
\tilde{r}^2}\bigr) \bigr]\label{f4} 
\end{eqnarray}
is the classical skyrmion mass, 
\begin{eqnarray}
a(F)={1\over e^3f_\pi }\tilde{a}(F)
=&&{1\over e^3f_\pi }\frac{8\pi }3\int 
{\rm d}\tilde{r}\tilde{r}^2\sin ^2F\bigl[ 1\nonumber\\
&&+F^{\prime 2}+\frac{\sin
^2F}{\tilde{r}^2}\bigr] \label{f5} 
\end{eqnarray}
is the classical inertial momenta and
\widetext
\begin{eqnarray}
\Delta M_j(F)=&&e^3f_\pi \cdot \Delta \tilde {M_j}(F)=e^3f_\pi \frac{-2\pi 
}{5\tilde{a}^2(F)}\int d\tilde{r}\tilde{r}^2\sin ^2F 
\Bigl[5+2(2j-1)(2j+3)\sin ^2F\nonumber\\
&&+[2j(j+1)+1]\frac{\sin ^2F}{\tilde 
r^2}
+[8j(j+1)-1]F^{\prime 2}-2(2j-1)(2j+3)F^{\prime 2}\sin 
^2F\Bigr]\label{f6}
\end{eqnarray}
is a negative quantum mass correction, $\tilde{r}=ef_\pi r$ being a 
dimensionless variable. The usual symmetric Weyl ordering for the operators 
are employed throughout (i.e. $\partial_0 G(q)=1/2\{\dot 
q^\alpha,\frac{\partial}{\partial q^\alpha}G(q)\}$). The operator ordering 
is thus fixed by the form of the Lagrangian~(\ref{f1}). No further 
ambiguity associated with the ordering appears at the level of the 
Hamiltonian~\cite{cebula92}. 

Minimizing Eq~(\ref{f4}) for $M(F)$, gives the usual classical solution 
$F(r)$ which behaves as $1/\tilde r^2$ at large distances. In semiclassical 
case~\cite{adkins83}, the quantum mass correction $\Delta M_j(F)$ is 
absent, and variation of Eq~(\ref{f3}) does not yield any 
stable solution~\cite{bander84,braaten85}. Such a 
semiclassical skyrmion~\cite{adkins83} was considered as 'rotating' 
rigid-body skyrmion with fixed $F(r)$. The canonical quantization procedure 
in collective coordinates approach leads to the energy~(\ref{f3}), 
variation of which produces a new integro-differential equation
\begin{eqnarray}
&&F^{\prime \prime }\Bigl[ -2\tilde r^2-4\sin ^2F+\frac{e^4\tilde 
r^2}{\tilde a(F)}\sin ^2F\left\{ 2R(F)-K(F)\right\} \Bigr] \nonumber\\
&&+F^{\prime 2}\Bigl[ -2\sin 2F+\frac{e^4\tilde r^2}{\tilde a(F)}\sin
2F\left\{ R(F)-K(F)\right\} \Bigr]\nonumber\\   
&&+F^{\prime }\Bigl[ -4\tilde r+\frac{e^4\tilde r^2}{\tilde a(F)}\sin
^2F\left\{ 4R(F)-2K(F)\right\} \Bigr]\nonumber\\
&&+\sin 2F\Biggl[ 2+2\frac{\sin ^2F}{\tilde r^2}-\frac{e^4\tilde r^2}{\tilde
a(F)}\Bigl\{ R(F)-\frac{8j(j+1)-6}{5\tilde a(F)}+2K(F)\Bigr\}   
\nonumber\\ 
&&-\frac{e^4}{\tilde a(F)}\sin ^2F\Bigr\{ 
2R(F)-\frac{12j(j+1)-4}{5\tilde a(F)}\Bigl\} \Biggr] =0\label{f7}
\end{eqnarray}
\narrowtext
where
\begin{equation}
R(F)=\frac 83\Delta \tilde {M_j}(F)+\frac{2\ell (\ell +1)}{3\tilde 
a(F)}+\frac{8j(j+1)-1}{5\tilde a(F)}\label{f8}
\end{equation}
and
\begin{equation}
K(F)=\frac{4(2j-1)(2j+3)}{5\tilde a(F)}\sin ^2F\label{f9}
\end{equation}
with the boundary condition $F(0)=\pi $ and $F(\infty )=0$.
Contrary to the semiclassical case, the asymptotic equation for 
$F(r)$ becomes at large $r$
\begin{equation}
\tilde r^2F^{\prime \prime }+2\tilde rF^{\prime }-(2+m_\pi ^2\tilde r^2)F=0,
\label{f10} 
\end{equation}
with 
\begin{equation}
m_\pi ^2=-\frac{e^4}{3\tilde a(F)}\left\{ 8\Delta \tilde 
{M_j}(F)+\frac{2\ell (\ell +1)+3}{\tilde a(F)}\right\}.\label{f11}
\end{equation}
The corresponding solution has the form
\begin{equation}
F(\tilde r)=k\left( \frac{m_\pi }{\tilde r}+\frac 1{\tilde r^2}\right) \exp
(-m_\pi \tilde r).\label{f12} 
\end{equation}
The integrals (\ref{f4}), (\ref{f5}), (\ref{f6}) converge ensuring stability 
of a quantum skyrmion only for $m_\pi ^2>0$. The positive quantity $m_\pi $ 
can be interpreted as the mass of pion. The negative quantum correction 
$\Delta {M_j}(F)$ changes the asymptotical behaviour of $F(r)$ 
radically. The quantum chiral angle $F(r)$ falls off exponentially even in 
the chiral limit. We have calculated numerically the solution of equation 
(\ref{f7}) which depends on the representation $j$, spin-isospin $\ell $ and 
the parameter $e$ of the model.

To perform numerical calculation of observables, 
we have made use of explicit expressions ~\cite{acus96a} for 
nucleon mass $m_N$, isoscalar radius $r_0$ of nucleon, the axial coupling 
constant $g_A$ and the magnetic moments of proton and neutron $\mu _p$ 
and $\mu _n$, respectively. The parameters of the Skyrme model $f_\pi $ and 
$e$ can be evaluated using two arbitrary empirical values. In the 
table~(\ref{table1}) we adjust $f_\pi $ and $e$ to fit the isoscalar radius 
and axial coupling constant. The mass of nucleon and the magnetic moments 
decrease slowly with increasing dimension of the representation. On the 
contrary, the mass of pion increases rapidly. The empirical values of $m_N$ 
and $r_0$ are fixed in calculations of the table~(\ref{table2}). Here the 
magnetic moments are decreasing with the dimension of representation, but 
the quantities $g_A$ and $m_\pi $ are increasing. The representation $j$ 
plays a role of a new discrete 'phenomenological' parameter. The best 
agreement with experimental data is achieved for $j=1$ in both tables. The 
agreement is much better then that obtained in the previous semiclassical 
calculations.  

\acknowledgments 

The authors would like to thank Professor D.O.Riska for useful
discussions. The research of this publication was made possible in part by 
the Long-Term Research Grants NN {\bf LA5000} and {\bf LHU100 } from the 
International Science Foundation.

\narrowtext
\begin{table}
\caption{The predicted static nucleon observables for different 
representations with fixed empirical values of isoscalar 
radius and axial coupling constant}
\label{table1}
\begin{tabular}{cdddc} 
{} &\bbox{j=1/2}&$\bbox{j=1}$&$\bbox{j=3/2}$&{\bf Exp.} \\ 
$m_N$&986.&948.&882.&$939$ MeV\\
$f_\pi $&61.4&58.9&55.7&$93$ MeV \\
$e$&4.37&4.13&3.92&  \\ 
$r_0$&input&input&input&$0.72$fm \\ 
$\mu_p$&2.70&2.53&2.42&$2.79$ \\
$\mu_n$&$-$2.14&$-$1.95&$-$1.84&$-1.91$ \\
$g_A$&input&input&input&$1.26$ \\
$m_\pi $&75.8&179.&259.&$138$MeV \\
\end{tabular}
\end{table}

\begin{table}
\caption{The predicted static nucleon observables for different 
representations with fixed empirical values of nucleon 
mass and isoscalar radius.}
\label{table2}
\begin{tabular}{cdddc} 
{} &\bbox{j=1/2}&$\bbox{j=1}$&$\bbox{j=3/2}$&{\bf Exp.} \\ 
$m_N$&input&input&input&$939$ MeV\\
$f_\pi $&59.8&58.5&57.7&$93$ MeV \\
$e$&4.46&4.15&3.86&  \\ 
$r_0$&input&input&input&$0.72$ fm \\ 
$\mu_p$&2.60&2.52&2.51&$2.79$ \\
$\mu_n$&$-$2.01&$-$1.93&$-$1.97&$-1.91$ \\
$g_A$&1.20&1.25&1.33&$1.26$ \\
$m_\pi $&79.5&180.&248.&$138$ MeV \\
\end{tabular}
\end{table}


\begin{references}

\bibitem{skyrme61a}
T.H.R.~Skyrme, Proc. Roy. Soc. {\bf A260},  127  (1961).
\bibitem{adkins83}
G.S.~Adkins, C.R.~Nappi and E.~Witten, Nucl.~Phys. {\bf B228},  552  
(1983).
\bibitem{bander84}
M.~Bander and F.~Hayot, Phys.Rev. {\bf D30},  1837  (1984).
\bibitem{braaten85}
E.~Braaten and J.P.~Ralston, Phys. Rev. {\bf D31},  598  (1985).
\bibitem{norvaisas94}
{E.~Norvai\v sas and D.O.~Riska}, Physica~Scripta {\bf 50},  634  (1994).
\bibitem{acus96a}
{A.Acus, E.Norvai\v sas and D.O.~Riska}, Nucl. Phys. {\bf A}, (in press).
\bibitem{fujii87}
{K.~Fujii, A.~Kobushkin, K.~Sato and N.~Toyota}, Phys. Rev. {\bf D35}, 1896
(1987).
\bibitem{cebula92}
{D.P.~Cebula, A.~Klein and N.R.~Walet}, J.~Phys. {\bf G18}, 499 (1992).


\end{references}
\end{document}